# Intense keV isolated attosecond pulse generation by orthogonally polarized multicycle midinfrared two-color laser field

Guicun Li,[1,2] Yinghui Zheng,[1,*] Zhinan Zeng,[1,†] and Ruxin Li[1]

[1]*State Key Laboratory of High Field Laser Physics, Shanghai Institute of Optics and Fine Mechanics, Chinese Academy of Sciences, Shanghai 201800, China*

[2]*University of Chinese Academy of Sciences, Beijing 100049, China*

We theoretically investigate the generation of intense keV attosecond pulses in an orthogonally polarized multicycle midinfrared two-color laser field. It is demonstrated that multiple continuum-like humps, which have a spectral width of about twenty orders of harmonics and an intensity of about one order higher than adjacent normal harmonic peaks, are generated under proper two-color delays, owing to the reduction of the number of electron-ion recollisions and suppression of inter-half-cycle interference effect of multiple electron trajectories when the long wavelength midinfrared driving field is used. Using the semiclassical trajectory model, we have revealed the two-dimensional manipulation of the electron-ion recollision process, which agrees well with the time frequency analysis. By filtering these humps, intense isolated attosecond pulses are directly generated without any phase compensation. Our proposal provides a simple technique to generate intense isolated attosecond pulses with various central photon energies covering the multi-keV spectral regime by using multicycle driving pulses with high pump energy in experiment.

## 1. Introduction

Attosecond (as) pulse generation, especially isolated attosecond pulse (IAP) generation, has opened the route to the probe of the ultrafast processes and real-time observation of electron dynamics of molecules and atoms on an attosecond time scale [1-5]. As the most powerful approach to generate as pulses, high-order harmonic generation (HHG), a highly nonlinear interaction of intense light with matter, has been extensively investigated during the past few decades [6-8]. HHG can be intuitively interpreted by the three-step model [9]. The outermost electron will be first tunneling ionized near the peak of the driving field, then accelerated, and finally recombines with the nucleus, emitting high harmonic photons. For multicycle long pulses, the three-step process will be repeated in every half optical cycle (OC), thus attosecond pulse trains (APTs) are generated [10]. In order to generate IAP, the harmonic emission must be confined within one half cycle, leading to a single light burst. Present schemes for isolated as pulse generation basically involves carrier-envelope-phase stabilized few-cycle driving laser pulses [11], polarization gating technique [12], or multi-color laser waveform control [13-16]. The shortest IAP obtained in experiment so far is 67as covering 55-130eV [17].

To date, HHG experiments have been widely performed by Ti: sapphire lasers at a wavelength of 800nm, and phase-matching harmonics are limited to 150eV. But in practical applications such as biological imaging, the water window (280eV-540eV) or keV X-ray pulses are of our special interest. To extend photon energies to this spectral range, midinfrared (MIR) laser fields have been used [18-21], which is primarily due to the cutoff law: $\hbar\omega_c = I_p + 3.17U_p$ [9], where $I_p$ is the ionization potential of the target gas, $U_p = 9.33 \times 10^{-14} I_L[W/cm^2] \times (\lambda[\mu m])^2$ is the acquired pondermotive energy of a free electron in the laser field. With the rapid development of frequency down-conversion technique by optical parametric amplification (OPA) and optical parametric chirped pulse amplification (OPCPA), intense MIR pulses with the wavelength up to 3.9 micrometers have been generated for HHG [19,20]. Using longer wavelength driving laser pulses combined with the extension of phase-matching technique into the X-ray spectrum, bright coherent ultrahigh order harmonics in the water window and keV X-ray regime have been generated [18,21]. Besides, soft X-ray attosecond pulses covering the water window have been experimentally generated by 1.9-cycle 1850nm MIR pulses [22, 23] and 1.7 $\mu m$, 2-cycle MIR polarization gating (PG) method [24]. However, the obtained harmonic yields are very low and attosecond pulses are very weak because of the unfavorable scaling of the single-atom response with laser wavelength [25, 26]. In order to increase the efficiency, multi-color waveform synthesizing has been extensively used to loose the requirement of driving pulse duration for IAP generation, decrease the pulse width of IAPs and enhance harmonic yields. Previous studies focus on parallel multi-color schemes for IAP generation [13-16, 27-29]. However, by adding an orthogonally polarized second pulse, the recollisions of free electrons can be manipulated two-dimensionally (2D), and quantum trajectories will be preferentially selected [30-33], thus an IAP can still be generated [34, 35]. In Ref. [34], the authors have theoretically investigated IAP in an orthogonally polarized two-color (OTC) laser field, but they still used few-cycle pulses, which is quite difficult and will limit the pump energy in experiment. In Ref. [35], the author has extended the driving pulse duration to more than ten OC for IAP by superposing an orthogonally polarized multicycle chirped laser upon the driving pulse, and a single 410as pulse can be obtained without any phase compensation, but this experimental alignment may be complicated, and the efficiency is low.

In this work, the generation of high-order harmonics and IAPs in an orthogonally polarized two-color laser field (OTC) consisting of a mid-infrared (MIR) fundamental pulse and its second harmonic pulse is investigated. Compared with traditional 800nm laser pulses, the cutoff can be easily extended to the X-ray spectrum by the MIR driving pulses. Besides, longer wavelength laser pulses can reduce the attosecond chirp because the chirp scales as $\lambda^{-1}$ [25], which is beneficial for short IAP generation. Moreover, our previous work showed that under the proper two-color delay, multiple electron trajectories can be reduced to few or even a single recollision [36], and the inter-half cycle interference effect becomes less pronounced even if driving pulses with longer pulse duration are used, due to the large phase difference of the harmonic emission that is proportional to the cube of the wavelength in each half cycle [36-38]. Based the facts above, in this paper, it is found that several quasi-continuum spectral humps that are stronger than other normal harmonics by about one order of magnitude, corresponding to intense IAPs as short as ~320as, are directly generated without any phase compensation in our MIR-OTC field, which is not feasible in 800/400nm OTC field. Unlike conventional supercontinua, the semiclassical trajectory model [9, 33] reveals that the quasi-continuum humps originate from the 2D control of electron-ion recollisions in our MIR-OTC field, thus providing a more advantageous technique to generate intense IAPs covering the water window or even multi-keV regime.

## 2. Numerical results and discussions

We simulate HHG in our MIR-OTC scheme by the strong-field approximation (SFA) model with the ground state depletion being considered [39]. The model atom in our simulation is helium (He). The OTC field can be expressed:

$$E_s(t) = E_x f(t)\cos(\omega t)\vec{x} + E_y f(t+\tau)\cos[2\omega(t+\tau)]\vec{y}, \qquad (1)$$

where $E_x$, $E_y$ represent the electric amplitudes of the fundamental and second harmonic pulses polarized along x and y directions, respectively. $f(t)$ is the Gaussian envelope, and $\tau$ is the two-color delay. According to Ref. [36], when the two-color delay is around zero, the number of recollisions will be greatly reduced. Therefore, $\tau$ is optimized to be zero in our simulation. The laser intensities of the fundamental and SH pulses are $6\times10^{14}$ W/cm$^2$ and $1.2\times10^{15}$ W/cm$^2$, respectively. Both the two pulses have the same duration (full width at half maximum, FWHM) of 8T (T is the optical cycle of the fundamental pulse, the same hereafter). By using these simulation parameters, we have investigated HHG in 800/400nm IR-OTC field and 1800/900nm MIR-OTC field, respectively, as demonstrated in Fig.1, where (a) and (c) represent the harmonic spectra obtained by the 800/400nm OTC field and 1800/900nm OTC field, respectively, (b) and (d) are the corresponding time frequency analyses of (a) and (c). It can be clearly seen from Fig.1 (a) and (c) that the cutoff is significantly extended when the 1800/900nm MIR-OTC scheme is used. Besides, the harmonic spectrum is less modulated than that by the 800/400nm OTC field. More surprisingly, multiple continuum-like spectral humps are generated near the cutoff. Each hump consists of about twenty orders of harmonics and its intensity is about one order higher than its neighboring normal harmonic peaks, which corresponds to an intense IAP in the time domain, as will be shown in Fig.3 (b).

According to Refs. [36-38], the harmonic spectrum can be interpreted as the interference effect of different quantum paths. Practically, only the trajectories whose dipole phase differences in each half cycle are not significant (usually $<2\pi$) can survive for the constructive interference and contribute to high harmonic emissions [40]. For multicycle OTC field, the phase difference of different trajectories for harmonic emission is proportional to $-\frac{U_p}{\omega} \propto \lambda^3$. Therefore, for the 800/400nm OTC scheme, the phase difference for each half cycle is very small, and many electron trajectories interfere constructively to contribute to HHG. This may be demonstrated in Fig.1 (b), where we can clearly see that there exist strong high harmonic emissions in every half cycle, leading to constructive inter-half-cycle interference effect of many electron trajectories and well-resolved harmonic spectra. Conversely, in the 1800/900nm MIR-OTC field, longer wavelength pulses will cause large phase difference of different half cycles, which will weaken the interference effect of different half cycles, thus leading to less pronounced spectral modulation. This case can be illustrated by Fig.1 (d), where we can clearly see that intense continuum-like humps occur corresponding to the cutoff region of each half cycle, as denoted by the red circles (labeled as 1~4). For example, for the first hump that covers 570th-590th (390eV~406eV) harmonics, the strongest harmonic emission occurs at the return time -2.685T. Although there exist three other recombination times at -2.135T, -1.625T and -1.115T contributing to harmonic generation, they are too weak and have little influence on the interference effect of different quantum paths, thus leading to a quasi-continuum spectrum that has less pronounced spectral modulation, as has been discussed above.

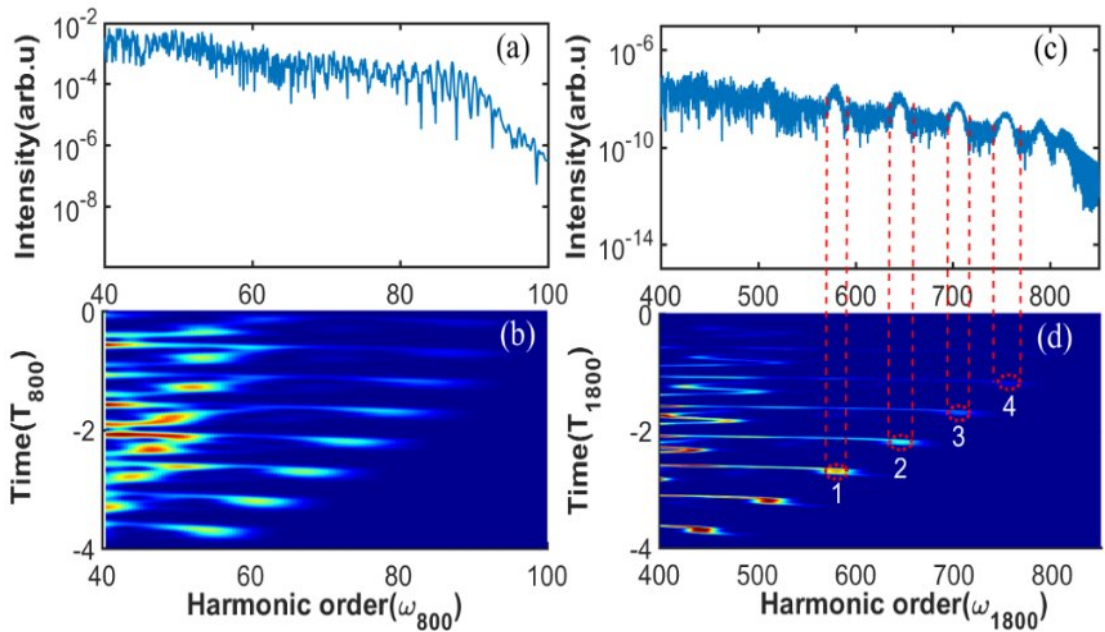

FIG. 1. (a) The harmonic spectrum obtained by the IR 800/400nm OTC field in the helium (He) atom. (b)The time frequency analysis of (a). (c) and (d), the same as (a) and (b), respectively, but for the MIR 1800/900nm OTC field. In both cases, the laser intensities are $I_\omega = 6 \times 10^{14} \, W/cm^2$ (1800nm or 800nm), $I_{2\omega} = 1.2 \times 10^{15} \, W/cm^2$ (900nm or 400nm), the two-color delay is fixed at zero, and the fundamental and SH pulses have the same duration of 8T. Note that diagrams (b) and (d) are plotted on the logarithmic scale.

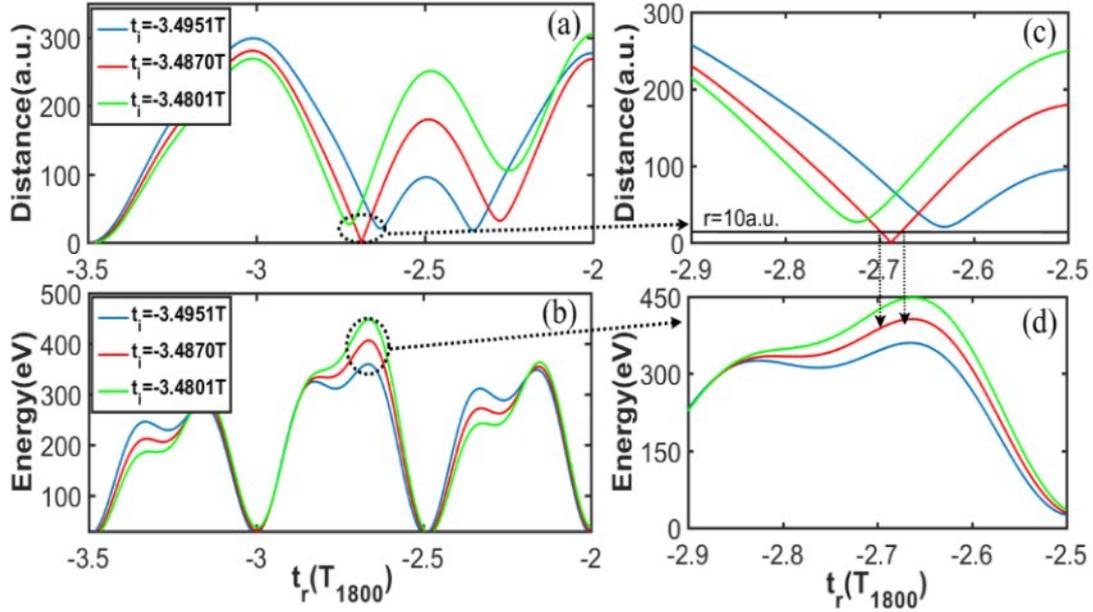

FIG. 2. (a) The radial distance from the origin of the coordinate system as a function of the time after the electron is freed by the 1800/900nm MIR-OTC field at -3.4951T (blue line), -3.4870T (red line) and -3.4801T (green line), respectively. (b) The photon energy of emitting high harmonics corresponding to (a) as a function of different instants after ionization. (c) and (d), enlarged diagrams of (a) and (b) in the time range of -2.9T~-2.5T, respectively. The solid black line in (c) indicates only the electron within 10a.u. from the ionic core can emit strong high harmonics for a specific spectral range, as denoted by the arrows in (d). Laser parameters are the same as those in Fig. 1 (c) and (d).

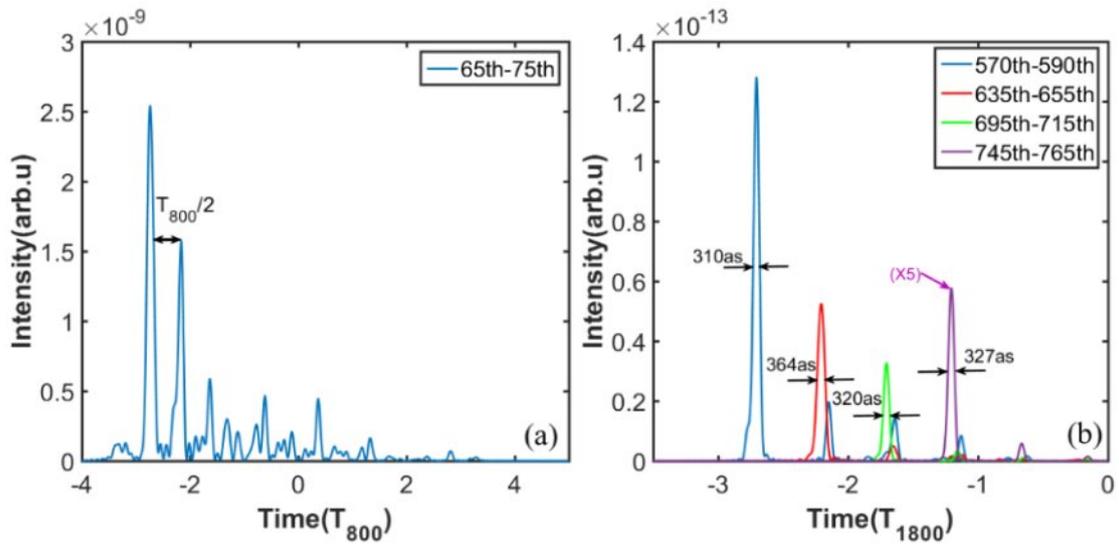

FIG. 3. The time profiles of generated sub-cycle pulses by superposing the harmonics (a) from the 65th to the 75th order in the 800/400nm OTC field (b) from the corresponding spectral humps in the 1800/900nm OTC field, respectively.

To get a more clear insight of the underlying dynamics of the electron trajectories contributing to these humps, we have also performed the semiclassical trajectory analysis in the 1800/900nm MIR-OTC field [9]. Electrons are freed at the origin position with zero initial velocity at different ionization instants, and the two-dimensional electron trajectories will be calculated by Newton's equation along the x and y directions, respectively [33]. It should be noticed that in the OTC field, the free electron does not necessarily need to exactly come back to the initial origin for HHG because of the quantum diffusion effect of the wave packet [39]. The returning wave packet can be assumed Gaussian and approximately expressed as:

$$\varphi_d(r,t) = (\pi\sigma_t^2)^{-\frac{1}{4}} \exp[-\frac{(r-r_c)^2}{2\sigma_t^2}], \quad (2)$$

where $\sigma_t = [\sigma_0^2 + (\frac{t}{\sigma_0})^2]^{\frac{1}{2}}$ and $r_c$ are the width and the center position of the diffused returning wave packet after the excursion time $t$, respectively. $\sigma_0$ 4a.u. is the initial wave packet width [41]. The recombination probability is proportional to the spatial overlap between the diffused returning wave packet and the ground state wave packet, and given by:

$$P_R \propto \int \varphi_d^*(r,t_r) \varphi_b(r) dr. \quad (3)$$

In the long wavelength MIR-OTC driving filed, the returning wave packet will diffuse seriously, and its width is much larger than that of the ground state wave packet $\varphi_b(r)$. Therefore, $\varphi_b(r)$ can be regarded as a delta function, and the integral of Eq. (3) is simplified as:

$$P_R = \varphi_d(r=0,t_r) = \sigma_t^{-\frac{1}{2}} \exp[-\frac{r_c^2}{2\sigma_t^2}]. \quad (4)$$

The harmonic yield is proportional to $|P_R|^2$, demonstrating that the HH yield decreases exponentially as the distance from the origin increases. Based on these discussions, we have calculated the electron trajectories at three different ionization instants around the electric field peak t=-3.5T, as shown in Fig. 2(a). The corresponding emitting photon energies are shown in Fig. 2(b). We only consider the return time with the half cycle -3T~-2.5T. This is because in -3.5T~-3T, the electron is just released and the wave packet width is smaller than the distance from the origin. According to Eq. (4), $P_R$ is very low. In -2.5T~-2T, however, the electron has a very long time to spread, and the width is very large, hence the amplitude of the wave packet is very small, leading to very low recombination probability as well. From Fig. 2(d), we can see the electron ionized at -3.4951T gains kinetic energy and emits high harmonics with the highest photon energy ~360eV (blue line), thus making no contrition to the emission of the first hump 570th-590th (390eV~406eV). However, when the electron is released at the later instant -3.4870T, it will return with the shortest distance (~0.2a.u.) from the initial origin. We set a criterion that only the electron within 10a.u. from the parent core has a substantial recombination probability to emit high harmonics [30], as denoted by the solid line in Fig.2 (c). This will lead to strong harmonic emission with photon energies 388eV~406eV (red line), just corresponding to the first hump 570th-590th (390eV~406eV), as indicated by the arrows in Fig.2 (d). In addition, the return time when the electron is closest to the origin position is -2.688T, well consistent with the emission time -2.685T obtained by the time frequency analysis in Fig. 1(d). Further, if the electron is freed at -3.4801T, the electron will emit higher photon energies exceeding the first hump (green line). Although the electron wave packet has a probability to recombine with the ground state wave packet and emits harmonics of the first hump, the distance from the origin is ~30a.u. from Fig.2 (c), much larger than that of the electron released at -3.4870T. In this case, the emission efficiency will be low. Therefore, the electron ionized around -3.4870T has the maximum recombination probability and makes the largest contribution to emit high harmonics corresponding to the first hump, while adjacent harmonics originate from the returns of the electrons that have a larger displacement from the ionic core, thus leading a much weaker harmonic emission. This may explain why the each hump near the cutoff in each half cycle is stronger than its neighboring normal harmonics. Similarly, other humps are generated in corresponding half cycles in the same way. By selecting the appropriate hump, an isolated attosecond pulse radiation will be generated. The detailed results are shown in Fig.3. It can be clearly seen from Fig. 3 (a) that by superposing the harmonics from 65th to 75th in the 800/400nm OTC field, an attosecond pulse with multi-peaks separated by half optical cycle is generated. However, by selecting the appropriate hump in the 1800/900nm MIR-OTC field, a clean ~320as pulse with a contrast ratio more than 9:1 is directly generated without any additional phase compensation, as shown in Fig. 3 (b), verifying that MIR-OTC driving laser pulses are more beneficial for IAP generation driven by multicycle pulses.

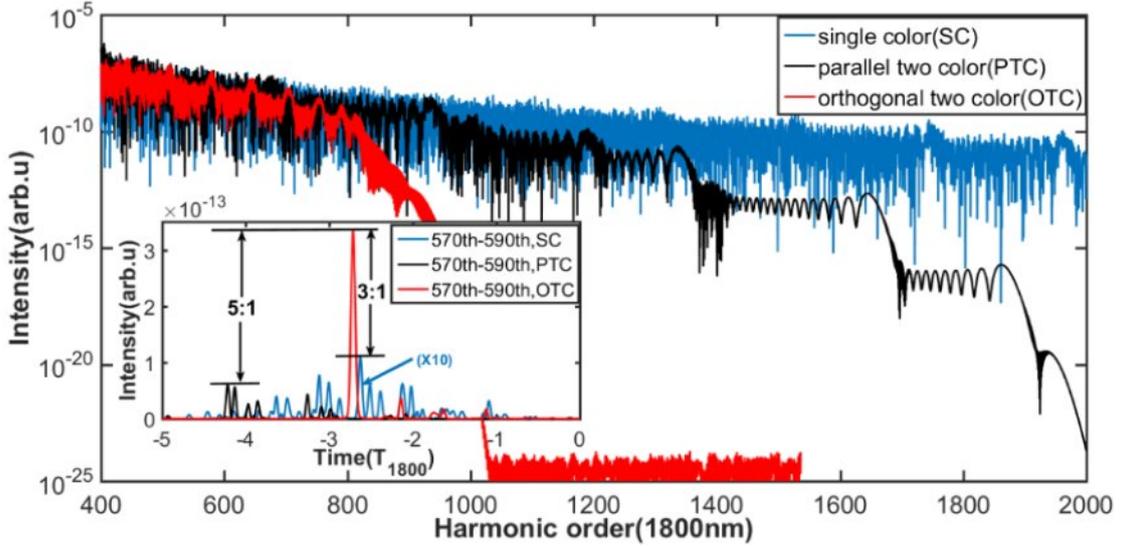

FIG. 4. The harmonic spectra obtained by the single-color (blue line), parallel two-color (black line) and orthogonal two-color (red line) fields, respectively. In both two-color schemes, the laser intensities are $I_\omega = 6 \times 10^{14} \, W/cm^2$ (1800nm), $I_{2\omega} = 1.2 \times 10^{15} \, W/cm^2$ (900nm). Both the fundamental and SH pulses are Gaussian shape with the same pulse duration of 8T, and the two-color delay is fixed at zero. For the single-color 1800nm field, the laser intensity is the sum of two-color schemes, $I_\omega = 1.8 \times 10^{15} \, W/cm^2$. The inset shows the time profiles by superposing 570th-590th harmonics (corresponding to the first hump in OTC field) generated by the three schemes.

Moreover, we have compared the results obtained by our MIR-OTC field with that by the widely-used parallel two-color field, and single-color field, as shown in Fig.4. In both two-color schemes (parallel and orthogonal), simulation parameters are the same as those in Fig. 1 (c) and (d). For the single-color 1800nm field, the laser intensity is the sum of two-color scheme, i.e., $I_\omega = 1.8 \times 10^{15} \, W/cm^2$. We can clearly see from Fig.4 that the cutoff is significantly reduced in the MIR-OTC scheme, compared with that by the single-color and parallel two-color fields, which is consistent with previous results [32]. Furthermore, for the single-color 1800nm field, the whole spectral region exhibits a severe modulation, indicating no IAP generation. For the parallel two-color field, a multiplateau structure is generated, which may be resulted from the deep ionization of the ground state population [42]. For the OTC field, however, the intensity of the humps is about five times stronger than that in the parallel two-color field, which is consistent with the results in Ref. [31]. But Ref. [31] revealed that only the low-order harmonics in plateau were enhanced. In our work, however, we find that the harmonic yield of the humps near the cutoff can also be enhanced. The corresponding time profiles by superposing 570th-590th harmonics (first hump in the OTC field) in three schemes are shown in the inset in Fig. 4, obviously demonstrating that only in the OTC field can a clean intense IAP be generated. Additionally, the intense IAP generated in our MIR-OTC scheme is about 5 times stronger than that in the parallel two color field and 30 times stronger than that in the single color field. It is worth pointing out that although IAPs can be generated by selecting the multiplateau supercontinua in the parallel two-color field, the efficiency is five orders lower than the humps.

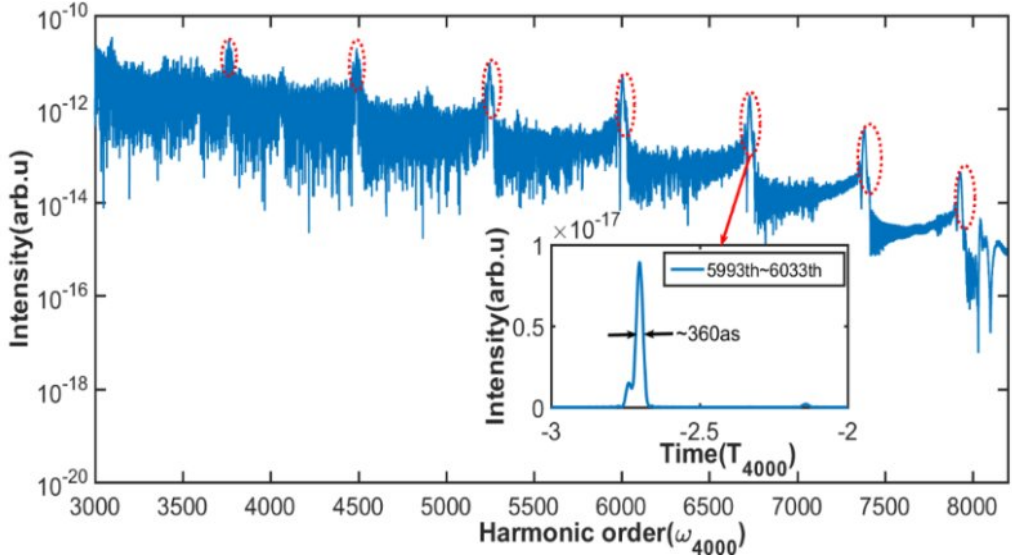

FIG. 5. High harmonic spectrum in 4000/2000nm OTC field. Humps with different central photon energies are denoted by red ellipses. Other simulation parameters are the same as those in Fig. 1. The inset shows the time profile by superposing H5993-6033 harmonics.

Finally, we discuss the feasibility and generality of this MIR-OTC scheme. Firstly, with the rapid development of OPCPA technique, intense MIR pump pulses with wavelength up to ~4 $\mu m$ are available in laboratory [19, 20]. Secondly, our MIR-OTC scheme can be directly applied for multicycle driving pulses, which will greatly relax the requirement for IAP generation and allow high energy pump lasers. Thirdly, the optimum simulation parameters discussed in our work, such as two-color delay and driving pulse duration, are general for the MIR-OTC scheme. In our $\omega+2\omega$ OTC scheme, the jitter between the two pulses can be neglected because the fundamental and SH pulses go through the same optical path, thus making the experimental setup more reliable and controllable. Another important issue is whether the humps still exist after considering the propagation effect. Our proposal is for intense IAP generation by multicycle high power laser sources which will be very loosely focused in HHG experiment, thus the macroscopic effect can be significantly eliminated [42, 43]. In such case, the single-atom hump structure still applies in experimental conditions. Based on this, we directly extend the MIR-OTC scheme to 4000/2000nm, which is more effective in experiment for high photon energy attosecond pulse generation. Figure 5 shows the single-atom harmonic spectrum from helium in the 4000/2000nm MIR-OTC field. Other simulation parameters are the same as those in Fig .1. It is obvious that the multiple humps with different central photon energies are more clear and distinguishable. The inset in Fig. 5 shows an isolated 360as pulse can be directly generated by superposing 5993th~6033th harmonics, covering 1857eV~1870eV photon energy region, without any additional phase compensation. The results show the possibility of generating intense IAPs covering the multi-keV spectral regime by the MIR-OTC scheme.

## 3. Conclusion and outlooks

In conclusion, the generation of high-order harmonics and IAPs in the MIR-OTC field has been specifically investigated with the semiclassical model and SFA model. By optimizing the two-color delay, it is found that multiple humps with different central photon energies are generated in the 1800/900nm OTC field, corresponding to clean and intense IAPs as short as 320as, which can be attributed to the 2D manipulation of the electron-ion recollision process and the suppression of inter-half-cycle interference effect in the MIR-OTC field. In particular, the humps generated by the MIR-OTC field are stronger than their adjacent harmonics by about one order of magnitude. Our MIR-OTC scheme, combined with phase-matching that has been extended to the soft X-ray region, will greatly relax the requirement of driving pulse duration for IAP generation and significantly enhance the attosecond pulse intensity at the same time. By optimizing the MIR-OTC scheme and choosing appropriate gas media, intense isolated attosecond pulses covering the multi-keV spectral range can be directly obtained.

**Acknowledgment**
This work was supported by the National Natural Science Foundation of China (Grants No. 11127901, No. 61521093, No.11134010, No.11227902, No.11574332, No.1151101142, No. 61690223 and No.11274325), the Strategic Priority Research Program of the Chinese Academy of Sciences (Grant No. XDB16) and the Youth Innovation Promotion Association of Chinese Academy of Sciences.

*yhzheng@siom.ac.cn
†zhinan_zeng@mail.siom.ac.cn